\documentclass[preprint]{aastex}

\shortauthors{LUFKIN, SARAZIN AND WHITE}
\shorttitle{TIME-DEPENDENCE IN CLUSTER COOLING FLOWS}

\slugcomment{Astrophysical Journal, in press}

\newcommand{\cool}{\rm cool}
\newcommand{\den}{{\rm\, cm^{-3} }}
\newcommand{\ergs}{{\rm \, erg\, s^{-1} }}

\newcommand{\Gyr}{{\rm\, Gyr}}
\newcommand{\K}{{\rm\, K}}
\newcommand{\kms}{{\rm\, km\, s}^{-1}}
\newcommand{\kpc}{{\rm\, kpc}}
\newcommand{\lum}{{\rm \, erg\, s^{-1}\, cm^{-3} }}
\newcommand{\mdot}{{\rm \, M_\odot\, yr^{-1}}}
\newcommand{\Mpc}{{\rm\, Mpc}}
\newcommand{\Msolar}{{\rm\, M}_{\odot}}
\newcommand{\yr}{{\rm\, yr}}

\begin{document}

\title{Time-Dependence of the Mass Accretion Rate \\
in Cluster Cooling Flows}

\author{
Eric A. Lufkin\altaffilmark{1,2,3},
Craig L. Sarazin\altaffilmark{1}
and
Raymond E. White III\altaffilmark{2,3}}

\altaffiltext{1}{Department of Astronomy, University of Virginia,
PO Box 3818, Charlottesville, VA 22903-0818}
\altaffiltext{2}{Department of Physics \& Astronomy, University of Alabama,
Box 870324, Tuscaloosa, AL 35487-0324}
\altaffiltext{3}{Code 662, NASA Goddard Space Flight Center, Greenbelt, MD 20771}

\begin{abstract}
We analyze two time-dependent cluster cooling flow models in spherical 
symmetry.  The first assumes that the intracluster gas resides in a static
external potential, and includes the effects of optically thin radiative 
cooling and mass deposition.  This corresponds to previous steady-state
cooling flow models calculated by White \&
Sarazin (1987).  Detailed agreement is found between steady-state models and 
time-dependent models at fixed times in the simulations.
The mass accretion rate
$\dot M$ is found either to increase or remain nearly constant once flows
reach a steady state.  The time rate of change of $\dot M$ 
is strongly sensitive to the value of the mass deposition parameter $q$, but 
only mildly sensitive to the ratio $\beta$ of gravitational binding energy to 
gas temperature.  
We show that previous scaling arguments presented by Bertschinger (1988) and
White (1988) are valid only for mature cooling flows with weak mass deposition
($q \la 1$).
The second set of models includes the effects of a secularly deepening
cluster potential and secondary infall of gas from the Hubble flow.
We find that such heating effects do not prevent the flows from reaching
a steady state within an initial central cooling time.
\end{abstract}


\section{Introduction} \label{sec:intro}
X-ray and optical data
strongly suggest that cooling accretion is taking place in more 
than half of galaxy clusters
(for a review, see Fabian 1994).
In non-cooling flow clusters, X-ray surface brightness profiles obtained 
from imaging data are usually well modeled by a 
single-temperature gas in hydrostatic equilibrium with a 
background isothermal potential (Jones \& Forman 1984).
In contrast, cooling flow clusters exhibit significant excess
central emission 
compared to that derived from a fit to an isothermal.
This excess emission is thought to be due to gas losing 
its thermal energy and condensing out of the intracluster medium
(ICM) at rates 
that often exceed $100\mdot$.
Over a cluster lifetime, the central dominant galaxies in
these clusters may accrete up to 
$\sim 10^{12} \Msolar$ of cooled gas.
(We assume a Hubble constant of $50 \kms\Mpc^{-1}$ and that a
cluster
lifetime is comparable to a Hubble time.)

X-ray surface brightness profiles of cooling flow clusters can be used
to infer mass accretion rates $\dot M_{\rm surf}$
which increase roughly linearly with radius from the centers
of cooling flows
(e.g.,  Stewart et al.\  1984; Thomas, Fabian, \& Nulsen 1987).  
If the flows are steady, this in turn implies that 
matter must be cooling and condensing
out of the ICM over the full range of radii ($\ga 100\kpc$) for which 
cooling flow emission is observed.
If there were no distributed mass
deposition, then gas would be deposited only at the center.
The resulting X-ray surface brightness profiles would be
significantly more centrally peaked than those observed.

X-ray spectra of cooling flows imply cooling rates $\dot M_{\rm spec}$
within a factor of two of those
inferred
from the imaging data
(e.g., Canizares, Markert, \& Donahue 1988;
Allen et al.\ 2000). 
This result
is an independent confirmation of the accretion rates derived from
imaging data.
The dynamical accretion rate in a spherical system,
$\dot M = 4\pi r^2\rho u$
(where $\rho$ is the gas density and $u$ is the flow velocity),
is not observed directly.
Although the dynamical and X-ray-derived accretion rates should be the
same if the flow is in a steady state,
the radial velocities in the accretion flow
are typically on the order of tens
of kilometers per second, well below both the velocity dispersion
of a
single galaxy and the spectral resolution of current X-ray
instruments.

Although the final repository for the cooling gas remains ambiguous,
the evidence is quite
strong that it is cooling
and decoupling from the flow
(Fabian 1994; 
White, Jones, \& Forman 1997;
Peres et al.\ 1998;
Markevitch et al.\ 1998;
Allen et al.\ 2000;
White 2000).
Heating processes are unlikely
to balance cooling since their functional dependences on density and
temperature do not match those for radiative losses.  
Indeed, previous numerical investigations have shown that
alternatives to large cooling accretion rates, such
as heat conduction (Bregman \& David 1988; Meiksin 1988), supernova
heating, and drag heating by orbiting galaxies (Bregman \& David
1989),
are able to reproduce the data only for narrow ranges in the
respective free parameters, if at all.  Moreover, it is difficult
to imagine a process other than cooling that can power an excess
emission 
rate as high as $10^{44}\ergs$ without producing other noticeable
signatures.  Finally, most alternatives to cooling accretion
fail to account for observed soft X-ray lines.

Models that assume a steady accretion flow with mass deposition
can reproduce the observed X-ray features of cluster cooling flows
(e.g., White \& Sarazin 1987, hereafter WS; Fabian 1994).
WS calculated steady-state models with 
a simple mass deposition formation law,
but such models provide no information on the evolution of cooling flows.

In this paper we use a suite of time-dependent spherical models to
assess the evolution of cooling flows in static and evolving
gravitational potentials.
In addition to providing predictions of the spatial structure
and spectral properties of relatively relaxed cluster cooling flows, 
the spherical
model also has implications for their long term evolution.
The time evolution of the accretion rate
has consequences for the total amount
of accreted material in cooling flows, and it can provide a test
for the models when compared to X-ray observations of high-redshift clusters.
Two free parameters likely to affect the time-variation of
$\dot M$ in the spherical models are the mass deposition efficiency 
and the ratio $\beta$ of gravitational
binding energy to thermal energy in the gas.
The parameter $\beta$ determines the shape of the gas
distribution in clusters when the gas is in hydrostatic
equilibrium.

Self-similar models can offer insight into the evolutionary effects, and they
provide a useful testing ground for fully time-dependent calculations
(Chevalier 1987, 1988;
Bertschinger 1989;
Lufkin \& Hawley 1993).
However, they are limited in that they allow either a narrow range of initial
conditions or a limited number of physical processes.  As a result, such models
have not been successful in attaining detailed agreement with the observations.
Simple scaling arguments are potentially useful (Bertschinger 1988;
White 1988), and we review these in \S~2 below.
Our numerical simulations of cooling flow evolution in static
gravitational potentials are presented in \S~3.

As a final point of inquiry, we examine the consequences of continued
cluster evolution by investigating
the effects of a secularly deepening cluster potential
and continued accretion of gas from the Hubble flow.
We present a simple physical argument in \S~2.3 which shows that
adiabatic compression of gas in a deepening gravitational potential 
does not inhibit cooling flows.
Our numerical simulations of cooling flow evolution in an evolving
gravitational potential are described in \S~4 and compared to similar
work by Meiksin (1990, hereafter M90).
Our conclusions differ from those in M90, which found 
that cooling flows are slow to reach steady state and
mass dropout rates were strongly reduced by
time-dependent effects in an evolving gravitational potential.
We do not find these effects in our simulations of cooling flows evolving
in the same deepening gravitational potential as used in M90,
which is consistent with the simple arguments we present in \S~2.3.
Our results are summarized in \S~5.

\section{Physical Arguments} \label{sec:physics}

Before proceeding to the numerical models, we first discuss some
physical arguments regarding the temporal behavior of cooling flow
clusters.  These include definitions of the various physical regions
in the flow, a review of analytic predictions of the time dependence 
of the cooling accretion rate $\dot M$,
and a discussion of whether adiabatic compression can extend the 
cooling time.  These physical arguments form a basis for the
discussion of the numerical models.

The most important process in models for
cooling flows is obviously the radiative cooling of the gas.
Over essentially all of the region of the flow, this occurs on a
timescale which is much longer than the dynamical time of the
gas, and as a result the flows and compression which occur are
relatively slow.
We do not include thermal conduction or other diffusive processes in
our calculations.
The only other processes which affect the temperature of the gas are
shocks and adiabatic compression or expansion of the gas.
In the absence of cluster mergers, the flow velocities are low and
shocks are not important in our models.
However, adiabatic compression of
the gas will occur as a result of the cooling and inflow of the
gas in the gravitational potential, or through the slow cosmological growth
in the cluster potential.
We discuss the effects of adiabatic heating on the cooling time scale of
the gas in \S~2.3 below.

\subsection{Timescales}

The general structure of cluster cooling flows in a static potential
is determined by
three characteristic timescales: the age of the cluster $t_{\rm age}$,
the cooling time $t_{\cool}$ and the dynamical time $t_{\rm dyn}$.
The instantaneous isobaric cooling time is
\begin{equation}
t_{\cool} = \frac{5}{2} \, { {kT} \over {\mu m_p \rho \Lambda(T)} } \approx
5 \times 10^9
\left( \frac{T}{10^8 \, {\rm K}} \right)
\left( \frac{\Lambda}{5 \times 10^{24} \, {\rm ergs \, cm^{3} \, s^{-1} \,
g^{-2}}} \right)^{-1}
\left( \frac{n}{10^{-2} \, {\rm cm^{-3}}} \right)^{-1} \yr ,
\end{equation}
where $\mu$ is the mean molecular weight
in units of the proton mass $m_p$ (we assume $\mu =
0.6$ in this work), and $n=\rho/\mu m_p$ is the total particle number
density.  The function $\Lambda(T)$ is the cooling rate such 
that $\rho^2\Lambda$ has the units $\lum$.
We assume the cooling function of WS
for half-solar abundances.  The dynamical time is
\begin{equation}
t_{\rm dyn} \approx 10^8 
  \Biggl({\sigma_c   \over 1000 \kms}\Biggr)^{-1}
  \Biggl({r_c\over 100  \kpc}\Biggr)\yr,
\end{equation}
where $r_c$ is the cluster core radius and $\sigma_c$ is the 
velocity dispersion of the cluster galaxies (also nearly equal
to the initial sound speed in the gas prior to cooling).

The cooling radius $r_{\cool}$ is defined as the point where the
instantaneous isobaric cooling time equals the system age.  Inside the
cooling radius,
the cooling time is less than the age of the system,
and a cooling flow occurs.  As long as the cooling time exceeds the
dynamical time, the gas flows subsonically into the center,
regulated by
cooling.
If the gas cools completely before reaching the center
and the rate of mass drop out is not too large ($q \la 3$),
the flow
generally passes through a sonic radius $r_s$, where the flow
speed equals
the local sound speed
(Sarazin \& Graney 1991).
For $r_{\cool}\gg r \gg r_s$ the
system is expected to be reasonably well-described 
as being in steady-state, an assumption which we test explicitly in
\S~3.3.
The validity of the steady-state approximation 
at $\sim r_{\cool}$ cannot be assessed except with time dependent models.

The above definition for the cooling radius is straightforward,
as well as familiar, but it
may not be the most useful.  Since the gas is flowing inward as it cools,
we consider whether a more relevant timescale may be the {\it integrated\/} 
isobaric cooling time
\begin{equation}
t_{\rm int} = {5\over 2P} \int_0^\theta 
  {\theta 'd\theta '\over \Lambda(\theta ')},
\end{equation}
where $\theta \equiv kT/\mu m_p$.
Typically, the integrated cooling time $t_{\rm int}$ is about a factor
of two shorter than the instantaneous cooling time $t_{\rm cool}$.
Consequently, gas may actually be taking part in the flow out to a considerably
larger radius $r_{\rm int}$, where $t_{\rm int}=t_{\rm age}$.
We comment more on this in \S~3.2.

\subsection{$\dot M$ as a Function of Time}

The time-dependent calculations of \S\S~3 and 4 solve explicitly
for the accretion rate $\dot M$ as a function of time.  
However, it may also be possible to 
estimate $\dot M(t)$ based on the assumption that
gas outside the cooling radius remains unaffected by cooling
in the interior.
Writing the dynamical accretion rate as $\dot M=4\pi r^2\rho u$, one
might expect that the velocity can be approximated by the
propagation rate of the cooling radius, so that $\dot M$ is interpreted
as the rate at which gas is swept over by the cooling radius.
This amounts to making the substitution
$u=dr_{\cool}/dt|_{t_{\cool}}$.
With this approximation, White (1988) showed that the cooling radius 
evolves with time as $r_{\cool} \propto t^{\eta}$,
where $\eta=[(1-\Delta_T\Lambda)\Delta_rT-\Delta_r\rho]^{-1}$,
and where we have used the notation $\Delta_x\equiv (d\ln/d\ln x)$.  
The exponent of the time dependence in the accretion rate $\dot M$
is then given by
\begin{equation}
\Delta_t\dot M = (3+\Delta_r\rho)\eta -1\equiv\xi.
\end{equation}
If the gas is initially isothermal, then $\xi < (>)~0$
when $\Delta_r\rho < (>) -3/2$.
In other words, cooling flow accretion rates will increase (decrease)
with time when the cooling radius is in a region where the gas density
profile is shallower (steeper) than $r^{-1.5}$.

The gas density profiles of clusters of galaxies are often fit
with the isothermal ``beta'' model.
In this model, the gas density distribution is given by
\begin{equation}
\rho (r) = \rho_0
\left[ 1 + \left( {r}\over{r_c} \right)^2
\right]^{- 3 \beta / 2} ,
\end{equation}
where $\rho_0$ is the central gas density
(e.g., Jones \& Forman 1984).
We will sometimes give $n_0 \equiv \rho_0 /\mu m_p$, which is the
corresponding total particle number density.
If the galaxies in a cluster are isothermal with a distribution
given by the analytic King approximation to an isothermal sphere,
then $\beta$ is determined by the ratio of the velocity dispersion
of the galaxies to that of the gas,
\begin{equation}
 \beta = {\mu m_p \sigma_c^2\over kT}.
\end{equation}
Typically, clusters have $\beta \approx 2/3$, and
the density profile tends to steepen
from $\Delta_r\rho\approx 0$ near the center to 
$\Delta_r\rho\approx -2$ at large radii.
Thus, there will be a transition radius $r_{\rm tran}$ where
$\xi\equiv \Delta_t\dot M$
changes sign from positive to negative.
For cases where $r_{\cool} > r_{\rm tran}$, one would infer from
equation (4) that $\dot M$ is decreasing.  
However, there appears to be some variation in the
value of $\beta$ from cluster to cluster, from 0.5 to about 1.2.
This variation translates into a variation in the gas scale height
relative to cluster core radii, and hence in the value of $\xi$ at
the cooling radius.
There is a correlation between cluster temperatures 
(both the gas temperatures at large radii and the dynamical
temperatures [velocity dispersions] of galaxies in clusters)
and the overall slope of the X-ray surface brightness profiles (White 1991), 
implying that hot clusters ($T_{\rm gas}\ga 7\times10^7\K$) tend to 
have density profiles scaling as $\rho\sim r^{-2}$ at large radii, while 
cool clusters ($T_{\rm gas}\la 5\times 10^7\K$) have shallower profiles.
There is therefore a range of cases, with both increasing and decreasing
mass accretion rates inferred via equation (4).  
In cases where imaging data imply a cooling flow, 
the slope of the X-ray surface brightness profile typically
does not make a sudden inflection at
either a core or cooling radius.  Consequently, determining
the transition radius from imaging alone is particularly difficult.  
Thus, in the absence of spatially resolved spectroscopy,
there is some {\it a priori\/} ambiguity in the
sign and magnitude of $d\dot M/dt$.  In \S~3, we
investigate this question for a plausible range in $\beta$.

\subsection{Adiabatic Compression}

One obvious limitation to the analysis of \S~2.2 is its failure to
account for ongoing dynamical evolution of the cluster itself.  
As subclusters merge, and as matter continues to accrete from the 
Hubble flow, the cooling flow may be altered, or disrupted altogether.  
A 1-D calculation can take these processes into account only
in the limit of quasi-static deepening of the cluster potential well.
Observations (Bird 1994) and numerical simulations (Evrard 1990) of 
hierarchical structure formation show that the growth of galaxy clusters
is only roughly approximated 
by a spherically symmetric deepening of a cluster's gravitational potential.
If the changes in the cluster potential result from violent subcluster 
mergers, shocks may be driven into the gas.
Such mergers cannot be easily modeled in a 1-D simulation.
Nonetheless, following work by M90, we consider a simple model
in which the growth of the cluster is described by a slow deepening
of the cluster potential and the resulting adiabatic compression of the
intracluster gas.
Although this is a simplistic approximation, we can still address
the question of whether such adiabatic compression is sufficient to 
change the qualitative nature 
of the steady flow compared to that of an isolated, static cluster,
as has been suggested by M90.

A simple physical
argument shows that adiabatic compressive heating is not sufficient to 
balance radiative cooling in intracluster gas.
The point is that, whatever the cause of such compression, it is a
reversible process and consequently does not change the entropy of the gas.
Thus, the only changes in the entropy are those due to radiative cooling.
Consider a parcel of gas initially at temperature $T_i$ and 
density $\rho_i$, with an adiabatic index $\gamma$.
For the purpose of this argument, we approximate the cooling
function as a power law, $\Lambda \propto T^\alpha$.
Adiabatic compression
to density $\rho_f$ will heat the gas to a temperature of
$T_f = T_i(\rho_f/\rho_i)^{\gamma-1}$.  
The cooling time scales as
$T^{1-\alpha}\rho^{-1}$, 
so, after compression, the cooling time is
\begin{equation}
t_{{\cool},f}
  =t_{{\cool},i}\left({\rho_f\over\rho_i}\right)
  ^{\gamma +\alpha -\gamma\alpha -2}.  
\end{equation}
For $\gamma=5/3$ the effect of compressional heating is exactly balanced
by the increased cooling rate for a critical value $\alpha =
\alpha^* = -1/2$.
We note that $\alpha\ga -1/2$ for almost all temperatures in the
range $10^6 \le T \le 10^9$, except for a narrow range
between $T\approx10^7\K$ and $T\approx3\times10^7\K$
(e.g., Raymond, Cox, \& Smith 1976; WS).
Thus, adiabatic compression generally should {\it decrease\/} the cooling
time for a given parcel of gas.  This result is independent of
the physical processes responsible for the compression.

\section{Time-Dependent Flow in a Static External Potential} \label{sec:static}

\subsection{Assumptions and Equations}

The models in this section are designed to resemble those of WS for
the purpose of assessing time-dependent effects in existing models
with mass deposition.  We therefore include 
radiative cooling and a mass deposition term,
but ignore the effects of conduction and
external heating, referring discussion of the latter
to the published literature.  Neglecting angular momentum, the 
corresponding spherically symmetric fluid equations are
\begin{eqnarray}
& & {d\rho \over dt} +{\rho\over r^{2}}{\partial\over\partial r}
(r^{2}u) = -\dot\rho \\
& & {du\over dt} +{1\over\rho}{\partial P\over\partial r}
    +{\partial\Phi\over\partial r }=0 \\
& & P{d\over dt}\ln (P\rho^{-\gamma}) = -(\gamma -1)\rho^2\Lambda
(T) \, ,
\end{eqnarray}
where $P$ and $T$ are the pressure and temperature
respectively, $\rho^2\Lambda$ is the cooling rate in $\lum$, 
$\Phi$ is the total gravitational potential and
\begin{displaymath}
{d\over dt} \equiv {\partial\over\partial t} +
u{\partial\over\partial r} \, .
\end{displaymath}
For the mass dropout rate
$\dot\rho$ we adopt the cooling-time law of WS, 
\begin{equation}
\dot\rho=q { \rho \over t_{\cool} }, 
\end{equation} 
where $q$ is an efficiency factor of order unity and $t_{\rm cool}$
is defined in equation (1).
This mass deposition law is based on the {\it ansatz\/} that
thermal instabilities due to cooling lead to growing perturbations which
leave the flow over a range of radii.
(It is often expedient in numerical simulations to cut off the
cooling function at some floor temperature $T_{\rm floor}$, the value of
which is arbitrary provided that the gas temperature does not reach
$T_{\rm floor}$ unless the flow becomes supersonic.) 

We solve equations (8)--(10) using 
a time-explicit hydrodynamics code in spherical 
symmetry (Lufkin \& Hawley 1993).  The code is similar to existing
second-order Eulerian codes except that it solves the fluid
equations in Lagrangian mode with a globally
conservative remap to the fixed grid.  The code has been tested on
a variety of problems, including self-similar cooling flows and cosmological
accretion (for details, see Lufkin \& Hawley 1993).

\subsubsection{The Assumed Mass Deposition Efficiency $q$}

All of the cooling flow models considered here have a finite 
mass deposition efficiency coefficient $q$.
If one assumes no mass deposition,
the mass accretion rate is constant with radius in a steady-state flow,  
and the gas cools catastrophically at the center (WS; Meiksin 1988, 1990), 
i.e., the cooling is sufficiently strong for the gas to pass
through a sonic point as it approaches the center.
The catastrophe is prevented (i.e. the flow remains subsonic
all the way to the center) for $q\ga 3.4$.  
The dynamical mass accretion
rate $\dot M$ is also found to be roughly proportional to radius for 
$q\approx3-4$.  Although a smaller $q$ cannot be ruled out 
observationally in all cases,
a value as small as $q = 0$ implies an X-ray surface brightness profile 
that is much too sharply peaked in the center.  
We therefore restrict our attention to the 
two cases $q=1$ (below which the central density is too sharply
peaked) and $q=4$, and examine 
the effect of this variation on model flows.

For these applications, the numerical grid covers the full dynamic range 
from scales of $\sim 1 \kpc$ near the center out to a radius of several Mpc, 
with a reflecting inner boundary 
($u\rightarrow0$ as $r\rightarrow0$).  
In the $q=1$ case there is generally a sonic radius
near $r\sim 1\kpc$ in the steady-state models.
The sonic radius is therefore unresolved 
by the grid spacing of $\sim 1\kpc$ near the center in the
time-dependent models.
If an outflow boundary inner condition is used, the grid must resolve the 
region between the sonic radius and the origin, in order to have 
supersonic outflow at the inner grid boundary (Lufkin \& Hawley 1993).  
Unfortunately, timestep constraints
prevent the use of an outflow boundary in this case,
owing to gas velocities in excess of $100\kms$ 
inward from the sonic radius.  

\subsubsection{The Background Cluster}

The initial conditions for the 
time-dependent simulations are chosen such that the gas is
isothermal
and in hydrostatic equilibrium with the external potential $\Phi$.
Integrating the hydrostatic equation we obtain for
the initial density profile 
\begin{equation}
 \rho(r)_{t=0}=\rho_0 \exp \left\{ - [\Phi(r)-\Phi_0] /
(kT_{\infty}/\mu m_p) \right\} \, ,
\end{equation}
where $T_{\infty}$ is the asymptotic gas temperature at large radii,
and $\rho_0$ and $\Phi_0$ are the density and potential at $r=0$.
The background potential for the cluster gas is assumed be that of
a massive central galaxy $\Phi_g$ plus a smooth cluster 
component $\Phi_c$.  The cluster mass distribution is 
taken as a King approximation to an isothermal sphere:
\begin{equation}
 \rho_c(r) =\rho_{c0} \left[ 1+(r/r_c)^2\right]^{-3/2} \, ,
\end{equation}
where $r_c$ is the cluster core radius.
The potential for this distribution is given by
\begin{equation}
 \Phi_c(r)=-\Phi_{c0}
  {{\ln\left[ r/r_c+\sqrt{1+(r/r_c)^2}\right]}\over{r/r_c}} \, .
\end{equation}
The central potential $\Phi_{c0} = 4\pi G\rho_{c0}r_c^2$
and is related to the line-of-sight velocity dispersion
$\sigma_c$ by
$\Phi_{c0} = 9\sigma_c^2$.
We assume 
$\sigma_c = 1054\kms$ ($\Phi_{c0}=10^{17} {\rm erg~g^{-1}}$) for the
hot cluster runs, and $\sigma_c=527\kms$ for the cool cluster runs.
The corresponding cluster masses inside of 250 kpc are 
$10^{14}$ and $5\times10^{13}\Msolar$, respectively.

We note that our adopted potential has a finite depth, whereas a true
isothermal is infinitely deep.  Consequently, the
density profile for the gas becomes sufficiently shallow at large
radii that the total X-ray luminosity diverges.  This actually follows
from assuming that the gas at large radii is hydrostatic and isothermal,
which cannot be true on spatial scales where the sound crossing time 
is greater than a Hubble time ($r\sim 10\Mpc$).  In the
evolving cluster models of \S~4, the outer parts of the cluster are in free
fall from the cold Hubble flow, with an accretion shock at a radius of
3--10 Mpc.
Because the cooling flow is confined to the inner 100
kpc or so, the solution is not sensitive to the structure of the
cluster much beyond this radius.  
We therefore quote total luminosities within one half and two Mpc.  
Furthermore, because gas densities far from the center 
are not well constrained observationally, we omit discussion of the
structure of the ICM at very large radii.

\subsubsection{The Central Galaxy}

All of the models here and below assume the presence of a massive,
stationary central galaxy.
However, in choosing a form for the galactic potential we
are guided by the desire to extend these models to higher
dimensions,
with the possibility of allowing a galaxy to orbit in the cluster
(e.g., Lufkin, Balbus, \& Hawley 1995).
The primary considerations are computational efficiency
and accuracy when the galaxy's position
changes with time.  The potential should be smooth and fully
resolved, with $\nabla \Phi_g\rightarrow 0$ as $r\rightarrow 0$
to avoid grid noise.
It should also be expressible in terms of simple analytic
functions. 
A potential of the form of equation (14) would be one possibility.
However, the mass in this model diverges logarithmically,
and it is generally preferable to have a galaxy potential
corresponding
to a finite mass.
While it is tempting to truncate an analytic King model at some 
finite radius, this can generate anomalous
sound waves for an orbiting galaxy on a finite difference grid.

A simple form which satisfies our numerical
criteria is Plummer's model (Binney \& Tremaine 1987):
\begin{equation}
\Phi_g = \Phi_{g0} \,
\left( 1+r^2/r_g^2 \right)^{-1/2}
\, ,
\end{equation}
where $r_g$ is a characteristic radius.
As it stands, the density distribution which gives rise to
equation (15) does not approximate real galaxies well.
However, we have found that a superposition of suitably weighted
components can give a good approximation to a King model.  
For example, the addition of two potentials, one for the
center and one for an extended distribution, can be written as
\begin{equation}
\Phi_g = \Phi_{g0}
  \left[ {(1-y )\over (1+r^2/r_g^2)^{1/2} }
        +{   y  \over (1+r^2/r_t^2)^{1/2} }\right] \, ,
\end{equation}
where $0<y<1$ and $r_t$ is the characteristic radius
of the extended component.
We have chosen to specify the potential directly, rather 
than starting from the underlying density distribution.  This is
because the
gas interacts with the galaxy only through gradients in the
potential.
The analytic form we have chosen contains functions which can be
calculated 
at high speed, whereas transcendental functions (such as power
laws, 
logarithms and arctangents) can increase the 
run time significantly.  The square root function is an exception. 
It uses
a divide-and-average algorithm, which converges very fast.  
A tabulated function, although not particularly expensive, 
can be rather unwieldy, especially when a galaxy's position
changes with time.

At large radii, ($r \gg r_t$), the potential tends to that of a 
point mass, so that the total mass has the finite value
\begin{equation}
M_g= {[(1-y )r_g+y r_t]\Phi_{g0}\over G} \, .
\end{equation}
For $y =0.15$ and $r_t^2=40r_g^2$, 
the gradient of this potential closely approximates
that of an analytic King model truncated at $r=20r_g$.

We note that in equations (13) and (14), $r_c$ is a true core
radius
(i.e., it corresponds to the radius at which the surface density
falls to one half of its central value).
In the model galaxy chosen here, this
occurs at a radius of about $ 0.63 r_g$.  
Figure 1 shows $g=-\partial\Phi_g/\partial r$
for a King model in which the central potential is 
$\Phi_{g0}=7\sigma_g^2$ (solid line; see Binney \&
Tremaine 1987),
compared with truncated (short dash) and nontruncated (dotted) 
King approximations to an isothermal sphere.  These are plotted along 
with a model galaxy with a potential of the form of equation (16),
for $y=0.15,\, r_g=1.58r_c,\, r_t=10r_c$ (long dash).
The vertical scale is in arbitrary units, and can be set either by
the central potential or the total mass of the galaxy $M_g$.

In the simulations below, the core radii $r_c$ are 250 kpc
for the cluster and 4.41 kpc for the central galaxy ($r_g=7\kpc$).  
Assuming a total mass for the galaxy of $2\times10^{12}\Msolar$ implies 
a line-of-sight velocity dispersion of $288 \kms$ at the center and 
$251\kms$ averaged over a circular aperture subtending 
a radius of $10\kpc$, in the manner of Bailey \& MacDonald (1981).  
The composite density distribution can also be reasonably
approximated by an NFW profile (Navarro, Frenk, \& White 1997)
with a scaling radius of $r_s\sim200$ kpc.

\subsection{Results}

Table 1 lists the input parameters and some of the calculated global
characteristics for ten simulations.  Each of five model clusters are
evolved for $15\,\Gyr$ for $q=1$ and $q=4$.  The first model, (runs
1 and 2) is taken to be typical of a strong cooling flow in a hot 
cluster, and is discussed in detail in \S\S~3.2.1. 
The second model (runs 3 and 4) corresponds to a relatively
cool cluster and is discussed in \S~3.2.2.
The remaining three cluster models, also discussed in \S~3.2.2, are
used to explore a range of values in $\beta$.
The results of runs 1--4 are plotted in 
Figures 2 -- 9, and are compared to steady-state calculations in
\S~3.2.3.

The physical assumptions, as stated above, are identical to those 
in WS, with the exception of the form of the external potential.  
Self-gravity is neglected for the static models, but is included in
the evolving cluster models in \S ~4 below.
Because the cooling flow is
primarily pressure-driven, the inclusion of self-gravity has little effect.

\subsubsection{Hot Cluster Runs}

Shown in Figure 2 are the density, temperature, mass accretion rate
and cooling times as functions of radius at various times in the
evolution, for the hot cluster run with $q=1$.
The initial conditions, given in Table 1, are chosen so that the
models have reasonable global properties at $t=10^{10}\yr$ in the
evolution.
In discussing the properties of the models, we use an age of
$10^{10}$ yr as a fiducial time for comparison to observed
clusters.
The solid curves in the figures refer to this fiducial age.
More generally,
the dotted, short-dashed, long-dashed, solid and dot-dashed lines
correspond to ages of $0, 3, 6, 10 {\rm~and~} 15 \Gyr$,
respectively.
The curve at 15 Gyr (beyond the fiducial age) is included to show
that a steady state has been reached.

Figure 3 shows the time evolution of both the instantaneous and integrated
cooling radii, $r_{\rm cool}$ and $r_{\rm int}$.
We also give the dynamical mass accretion rate
$\dot M = - 4 \pi r^2 \rho u$ as determined at each of these cooling radii.  
The X-ray-derived mass accretion rate (or cooling rate)
$\dot M_X$ is determined by assuming that all of the X-ray luminosity from
within the cooling radius is due to gas cooling within that radius.
This is the standard assumption made in analyses of the X-ray surface
brightness profiles to derive the total cooling or accretion rates
$\dot M_{\rm surf}$
(e.g., Thomas et al.\ 1987).  
(Recall that $\dot M_X=\dot M$ only for steady state flows.)  
As expected, there is close agreement between $\dot M$ and $\dot M_X$ at each
radius, although the value of $\dot M$ at $r_{\rm int}$ is nearly twice that 
at $r_{\cool}$.  The panel at lower right shows the value of
$\xi\equiv \Delta_t\dot M$ as derived from local gradients in the 
simulation (eq.~4).
We find $\xi\approx 0.29$ at $r_{\rm int}$ and 
$\xi\approx 0.59$ at $r_{\cool}$.
The value at $r_{\cool}$ compares reasonably to a value of
$\Delta_t\dot M\approx0.34$ measured directly from the simulation
by least-squares fitting of a straight line to $\log\dot M(t)$
as measured at $r_{\rm cool}$ and dumped every $10^8$ years.

We determine the transition radius 
(where $\xi=0$ in eq.~[4]) by examining the numerical results directly:  
$r_{\rm tran}=234\kpc$ at $t=10^{10}\yr$. 
The measured cooling radius at this time is $r_{\cool}=100\kpc < r_{\rm tran}$,
so we would infer from equation (4) that the accretion rate is 
increasing, as Figure 3 confirms.  
There are several sources of error in the estimate of $r_{\cool}$.  
The grid spacing is about $5\kpc$ in this region, although this can
be reduced if necessary by running the problem at higher resolution.  More
serious is the assumption that the relevant characteristic
radius is $r_{\cool}$, not $r_{\rm int}$ (see eqs.~1 and 3).
At $t=10^{10}\yr$ in the $q=1$ case, the location
where $t_{\rm int}=t_{\rm age}$ is $r_{\rm int}=151\kpc < r_{\rm
tran}$.  Thus we see that the scaling relation of \S~2.2 
is confirmed in this case for both measures of the cooling radius.

Several other features are noticeable from the figures.  Gas in the
core has cooled completely between 6 and 10 Gyr, with little change
in the flow thereafter.  
The initial cooling time at the center is 10.8 Gyr,
but the integrated isobaric cooling time is 6.43 Gyr.  
This time period corresponds to the phase in which the cooling radius 
and mass accretion rates slow to a steady rate of increase (Figure 3).  
The particular shape of the temperature profile is determined by the 
cooling function and the shape of the potential, 
while the density and accretion rate profiles are nearly featureless.  
Because the sonic radius is unresolved, the innermost
grid zones reflect low-amplitude sound waves into the flow.  
These are visible in the accretion rate at 15 Gyr, but do not
significantly affect the flow.

One might expect an inflection point in the density profile as the
center begins to cool, but this is not observed.  The entire
inner region cools almost simultaneously, but isobarically,
resulting
in a smooth transition to the steeper density profile seen at later
times.

The hot cluster $q=4$ solution (run 2) is shown in Figures 4 and 5.
This model has a higher rate of mass drop out at larger radii.
In this case, the initial density required to recover an accretion
rate of about $300\mdot$
at $t=10\Gyr$ is sufficiently high ($n_0=0.033\den$)
that the center cools in about $5\Gyr$.  
After this time we find that $\dot M$ is slowly decreasing,
even though both cooling radii ($r_{\cool}=74\kpc;~r_{\rm
int}=128\kpc$)
lie well inside $r_{\rm tran}=354\kpc$.
Moreover, the logarithmic rates of change of $\dot M_X$
measured at both $r_{\rm int}$ and $r_{\rm cool}$ are approximately
equal $\xi \approx -0.45$, whereas the value predicted by equation (4)
goes no lower than $\xi=-0.2$ for this time frame.  
The value of $\eta\equiv d\ln r_{\cool}/d\ln t$
is also discordant; 
we measure $\eta\approx0.5$ from the simulations, whereas the density 
and temperature profiles would suggest a value of $\eta\approx1.0$.
Thus, it appears that the scaling argument detailed in \S ~2.2, 
which assumes $q=0$, breaks down for large values of $q$.  
We will comment more on this below.
Because there is no sonic point in this model, there is no need
for high resolution in the innermost regions of the flow.

\subsubsection{Cool Cluster Runs}

We perform a second pair of simulations in which all the input parameters
are the same as those above except that the cluster has a central
velocity dispersion of $\sigma_c=527\kms$ 
and the asymptotic gas temperature is $T_{\infty}=3\times10^7\K$.
These parameters imply $\beta=0.6$ (run 3 and 4 in Table 1),
compared with $\beta=0.75$ in the hot cluster simulations.
The results for runs with $q=1$ and $q=4$ are shown 
in Figures 6--9.  The same general comments regarding the hot cluster
models also apply in this case.  The $q=1$ case shows an accretion rate
$\dot M$ that is increasing with time, in agreement with the arguments 
in \S~2.2,  while the $q=4$ case shows a nearly constant $\dot M$ even 
though the predicted $\xi$ is clearly positive near the cooling radius 
and nowhere negative.

There is some variation in the particular values of $\xi$ and $\eta$
compared with the hot cluster runs.  To see if these quantities correlate
with the value of $\beta$ in these models, we ran a series of cool cluster
models in which only the initial temperature was allowed to vary.
All other external parameters are the same as those in runs 3 and 4.
These are listed as runs 5--10 in Table 1, and the derived values
of $\eta$ and $\xi$ (taken at $r_{\cool}$ at $t=10^{10}\yr$)
are plotted versus $\beta$ in Figure 10.
Other than the extreme case of $\beta=0.36$, (which is included only 
to show that we recover the expected limiting case 
$\xi ,\eta\rightarrow\infty$ as $\beta \rightarrow 0$),  
the variation with $\beta$ is slight compared
to the dependence of $\xi$ and $\eta$ on $q$.  In every case the values
of $\xi $ and $\eta$ derived by measuring the time derivatives directly
from the models are much closer to the expected values when $q=1$ than
when $q=4$.  We must therefore conclude that the scaling argument of
\S~2.2 applies only in the limit of small $q$.

The reason for the inability of equation (4) to predict the flow behavior
for large $q$ is not immediately clear, although there are several ways
in which the flow may be affected by a large value of $q$.  One is that
the rapid mass deposition can significantly reduce the gas density 
in the interior, thereby reducing the pressure, so that gas slumps
to
the center from radii well beyond the cooling radius. Unfortunately,
there does not appear to be a well-defined position at which to evaluate
$\xi$ from observed density and temperature profiles.  This is because $q$
is not known in real clusters, and it may be that the relevant cooling
radius should be redefined again to account for the high rate
of mass deposition in this model.
Another possible cause for concern is that if
the mass deposition formula (eq.~11) is applied throughout the flow,
it implies that matter can condense
out even if the local cooling time exceeds the age.  To guard against
this, we ran models 5--10 with a cutoff to the mass deposition 
($q\rightarrow0$) for $t_{\cool}>t_{\rm age}$.  
We also repeated the first four runs with
this modification, but there were no differences in the results other
than a leveling-off in the dynamical accretion rate at the cooling radius.

\subsection{Correspondence to Steady-State Models}

We have seen that the dynamical mass accretion rate is a good approximation 
to the accretion rate one would infer, under the assumption of steady flow,
from X-ray observations of the model clusters described above.
A further test of the assumption of steady flow
is to compute steady-state models corresponding
to a fixed time in the evolutionary models.  The equations for steady flow
are given by WS, and are obtained by setting all partial time derivatives
to zero in equations (8)--(10).  
The resulting ordinary differential equations are then solved numerically 
subject to appropriate boundary conditions.
Because the time-dependent calculation used here and the ODE solver
are independent, this is also a further test of the hydrodynamics code.

There are a number of possible ways to specify the three boundary conditions
on the steady-state models.  Following WS, we choose to fix the age
($t_{\rm age}=10^{10}\yr$), and values for the temperature
$T|_{r_{\cool}}$ and the dynamical mass accretion rate 
$\dot M|_{r_{\cool}}$ at the cooling radius.
For direct comparison with the time-dependent models, 
we can take the values from Table 1.  The solution to the
steady flow equations is plotted in Figures 11 (hot cluster) and 12
(cool cluster).  These curves can be compared directly to Figures 2 and 4
at $t=10\Gyr$ (solid lines).  The agreement is very
close, with only a small departure near $r=0$ in the models with $q=1$.
This difference is not unexpected given the artificial boundary conditions
$u\left|_{r=0}\right.=0$ imposed in the time-dependent 
calculations on flows with sonic radii.

\section{Time-Dependent Flow in a Static Evolving Potential}
\label{sec:evolving}

One possible limitation to the preceding models is the neglect of cluster
evolution.  This can consist of dynamical evolution in the form
of continued mergers or two-body encounters
which act to deepen the cluster potential, or of continued
infall of gas from the Hubble flow.
The gas itself may contribute to the change in the total potential.
M90 included all of these effects in a series of simulations in spherical 
symmetry.  
Two classes of models were considered: one with a centrally 
dominant galaxy (because all cooling flows have one) and one 
without a central galaxy.
We reexamine the effects of cluster evolution on cooling flows
and use the same cluster models as M90 so we may
compare results.
We are interested in whether 
the flows reach
a steady state with or without mass deposition.  

\subsection{Model Characteristics}

In the models of M90, the cluster potential is as described 
above in equation (14), 
with an initial velocity dispersion of about $500 \kms$.
The depth of the potential is assumed to increase
gradually with time until, at the end
of each run,  the velocity dispersion is about $1000 \kms$.  This is done
to mock the dynamical evolution of the cluster from a redshift of
$z=1.5$ to $z=0$, corresponding to a time span running from 3.3 Gyr
to 13.1
Gyr (for $\Omega_0=1$ and $H_0 = 50 {\rm\, km\, s}^{-1}{\rm\, Mpc}^{-1}$).
The gas was assumed to trace the dark matter initially and
the central gas temperature put as much energy per unit mass
in the gas as in the dark matter.  The initial temperature profile was
assumed to be adiabatic.

Models without a central galaxy assumed that the initial gas velocity 
follows the same profile as that for collisionless infall onto a region 
of overdensity equivalent to that given by the total 
cluster mass interior to each point.  
The gas is therefore at rest only
at $r=0$ and at the turnaround radius far outside the cluster, with infall 
interior to the turnaround radius.  Initially the gas in the cluster is 
thus not in hydrostatic equilibrium, but instead falls inward, then bounces
adiabatically.  The gas in the center relaxes on
a dynamical timescale, which is much less than the cooling time,
so the precise initial conditions do not seriously affect the
subsequent evolution of the flow.  M90 found that gas
will cool catastrophically at the center of such flows if $q=0$,
but that
for $q=1/2$ the flow is stabilized against runaway cooling.
(Note that M90 employs a different definition of $q$, and the
value of $q = 1/2$ in M90 corresponds to $q = 5/6$ in our notation.)

We are primarily interested in the models of M90 which contain a central
galaxy in the initial conditions.  In this case, 
the initial velocity profile is the same as in the case without a 
central galaxy except that $u=0$ for $r<250\kpc$, resulting in 
a discontinuous initial velocity profile.

\subsection{Numerical Simulations} 

Figure 13 shows our simulations using the PLPC code 
(Lufkin \& Hawley 1993) for the
fiducial model of M90
with a central galaxy and without mass dropout ($q=0$).
All of the parameters
given in M90, including the potential for the central galaxy
and the self gravity of the gas, are duplicated here.  (We use
a different cooling function, which we comment on below.)
Up until the second time frame, our results
and M90 agree in detail.  We note that the discontinuous velocity
profile results in an impulsive squeeze on the initially hydrostatic
atmosphere interior to $r_c$.  This strong pressure wave travels
into the center, bounces off the reflecting inner boundary, and travels 
back out into the infalling gas beyond $r_c$.  This is evident in 
both the velocity and temperature profiles at $t=4\Gyr$ 
(dashed line in Figure 13).  We find, using fine time resolution in
the data dumps, that two strong bounces occur before the gas relaxes
at the center.
(The relevant plots in M90 also show evidence for these bounces.)
The run begins at $t=3.3\Gyr$, when the central instantaneous
isobaric cooling time is $9.0\Gyr$.

For time slices later than 4.0 Gyr, our results differ from those in M90.  
We find that the gas cools in less than an initial cooling time,
with an ensuing cooling catastrophe at the center ($t=9.9\Gyr$,
dot--dash line of Figure 13).
For the same model, M90 found no cooling catastrophe.

We have made a number of tests to search of the source of this discrepancy
with the results of M90.
First, the calculations were repeated using exactly the same
grid of 150 zones as used in M90:
\begin{eqnarray*}
r_i = & 1.88 \kpc (i-1) \qquad & i   \leq   21 \\
r_i = & 1.046 r_{i-1}   \qquad \hfill &  i > 21 \, .
\end{eqnarray*}
For comparison, we also made plots which were 1-2-1 smoothed (as in M90),
but detected no variation from those shown in Figure 13.  
Second, we considered the effects of differences in the cooling function
in the two sets of models.
The cooling function adopted in M90 is that of Gaetz \& Salpeter (1983), 
whereas we have used that of WS, which is somewhat stronger.
We therefore ran the case with a central galaxy with
the strength of the cooling function reduced by 20\% to compensate.  
The result was virtually unchanged.
This is not surprising, since the cooling rate is proportional
to the square of the density.
It therefore appears unlikely that the disparate results are caused
by small differences in the cooling functions.
Finally, we have run the fiducial model with two other hydro codes,
the direct-Eulerian MONO scheme of Hawley, Smarr, \& Wilson (1984) and
a modified version of the implicit scheme of Ruppell \& Cloutman (1975).  
Although the exact time of cooling collapse varies by about 20\%
(this measure is very sensitive to the value of the central density,
which in turn is sensitive to the numerical diffusion as the impulsive
compression waves bounce off the center early in the simulation),
the density, velocity and temperature profiles
agree in detail with the calculation of Figure 13.

One possible reason for this discrepancy is the grid noise
(zone-to-zone variations in the fluid variables) present in
the numerical calculations of M90.  This could cause sound waves to be 
amplified or converted into heat energy, perhaps via artificial viscosity.  
We also note that the crests in the temperature
and velocity profiles are sharper in our calculation than in the
figures of M90.
This is true even when our model output is smoothed identically to
that present in M90.
This suggests that the features in the plots in M90 are spread over
several grid zones, which might indicate that a significant level of
numerical diffusion was present.
Such diffusion may have transported
heat energy into the core.

We find that a cooling catastrophe occurs either with or without 
the central galaxy.  We do reproduce the result in M90 that the catastrophe
occurs sooner in the case without a central galaxy even though the initial
cooling time is longer.  As suggested in M90, this is likely due to the 
homologous collapse of the constant-density core.  

When mass deposition is included, the flow reaches a steady state on an
initial cooling timescale.  The details in the flow variables appear to be
affected by transient features left over from the discontinuous initial 
conditions ($q=1$ and 4; Figures 14--17).  In particular, there 
are discontinuities in the position of the cooling radius due to
the shock waves that bounced off the center at the beginning of the
runs.
As expected, the temperature beyond 100 kpc increases gradually 
due to continued shock heating from secondary infall.  

The agreement between the cooling and dynamical accretion rates
is also quite close, as in the static models of \S~3.
Thus, we conclude that flows with mass deposition do reach a steady state
when cluster evolution and secondary infall are included in the spherical 
model.  For both runs, $\dot M$ stays constant to within a 
factor of two after the flows reach a steady state.
Evolutionary effects in spherical symmetry therefore do not
appear to affect the time evolution of the
mass accretion rate relative to that found for static clusters.  

\section{Discussion and Conclusions} \label{sec:conclude}

We have solved for the time-dependent behavior of cooling gas 
in a variety of spherically symmetric cluster models, with
particular emphasis on the evolution of the mass accretion rate.  

We find that the steady-state approximation is valid after the initial onset
of cooling at the center for cluster flows in both static and evolving
external potentials.  This result is insensitive to the inclusion
of self-gravity.  For the models considered here
the accretion rate either increases or stays about the same with time.  
While the rates of increase or decrease that we see in the simulations
would be difficult to infer from imaging observations, spatially resolved
spectra can be a useful diagnostic (e.g., Wise \& Sarazin 1993).
The difference arises from the fact that while the age is an assumed 
parameter in the models, it can be {\it measured\/} by calculating the
cooling time at the radius of extent of soft X-ray lines.  Moreover, the
cooling rates derived from imaging are more model-dependent, than those
derived from spectra, which can in principle be obtained from a measure of 
the mass cooling through a single line.

A decreasing accretion rate with time is seen only for
cases where the temperature of the gas corresponds
closely to the cluster velocity dispersion ($\beta=1$).  In a
number of clusters, especially poorer ones, the gas may be hotter by a factor
of two than the virial temperature measured from galaxy motions.  
Thus, although we cannot
rule out a decreasing mass accretion rate on the basis of these experiments,
nearly constant or increasing accretion rates are favored for most clusters,
unless they have been strongly affected by mergers.
In the absence of mergers,
the spherical model with mass deposition would predict that
cooling flows were not much more vigorous in the recent 
past than they are today.  
Attempts to include additional evolutionary effects via secular
deepening of the cluster potential well and continued accretion
from the Hubble flow in spherical symmetry do not alter these results.  
However, it is likely that cluster mergers can have a strong effect
on cooling flows
(e.g., McGlynn \& Fabian 1984;
G\'omez et al.\ 2000).
Further study of the effect of cluster evolution on cooling flows
will require numerical simulations in three dimensions
with high spatial resolution and including cooling.

\acknowledgements
We would like to thank Steve Balbus, Joel Bregman, John Hawley, Brian McNamara, 
Avery Meiksin, Bill Sparks and Mike Wise for numerous informative discussions.
E. A. L. acknowledges the support of NSF grants PHY90-18251 and AST-8919180.
C. L. S.  was supported in part by NASA Astrophysical Theory Program grant
NAGW--2376.
R. E. W. III was supported in part by the NSF and the State of Alabama
through EPSCoR II and by a National Research Council Senior Research
Associateship at NASA GSFC.

\clearpage

% Figure 3.
\figcaption[fig1.ps]{
Plot of gravitational acceleration $g$ versus radius for four
model galaxies.  The dotted line is a King approximation to
an isothermal sphere ($\rho\sim r^{-3/2}$ at large $r$).
The short-dashed line corresponds to the same distribution
truncated at 20
core radii.  The long-dashed line gives the acceleration due to the
superposition of two Plummer's models (derived from equation 16),
with the
parameters as stated in the text.  The solid line is for a true
King
model with the central potential equal to $7\sigma_c^2$.
}

% Figure 2.
\figcaption[fig2.ps]{
The fully time-dependent solution for the hot cluster $q=1$ model
(run 1 in Table 1).
The four panels show total number density, temperature,
dynamical mass accretion rate, and instantaneous isobaric cooling
time
as functions of radius shown at
$t=0$ (dotted line), 3 (short dash),
6 (long dash), 10 (solid) and $15\Gyr$ (dot-dash).
}

% Figure 3.
\figcaption[fig3.ps]{
Comparison of accretion rates for the $q=1$ model of Figure 2.
The top left panel shows the time evolution of the cooling radius
determined from the instantaneous isobaric cooling time (eq.~1, solid
line) and
the integrated isobaric cooling time (eq.~3; dotted line).
The upper right and lower left panels show the dynamical mass
accretion
rate $\dot M=-4\pi r^2\rho u$ and the cooling mass
accretion rate derived from the X-ray emission $\dot M_X$, respectively.
The solid (dotted) curve corresponds to taking the accretion rate
at
the radius where the instantaneous (integrated) cooling time equals
the
age.  The panel at lower right gives the estimated logarithmic time
derivative
of the mass accretion rate as a function of cooling radius
(eq.~4) at the end of the run.
}

% Figure 4.
\figcaption[fig4.ps]{
The time-dependent solution for the
hot cluster $q=4$ model (run 2 in Table 1).
The notation is the same as Figure 2.
}

% Figure 5.
\figcaption[fig5.ps]{
Same as Figure 3, but for the hot cluster $q=4$ model (run 2 in Table 1).
}

% Figure 6.
\figcaption[fig6.ps]{
Cool cluster model with $q=1$ (run 3 in Table 1).
The notation is the same as in Figure 2.
}

% Figure 7.
\figcaption[fig7.ps]{
Cool cluster model with $q=1$ (run 3 in Table 1).
The notation is the same as in Figure 3.
}

% Figure 8.
\figcaption[fig8.ps]{
Cool cluster model with $q=4$ (run 4 in Table 1).
The notation is the same as in Figure 2.
}

% Figure 9.
\figcaption[fig9.ps]{
Cool cluster model with $q=4$ (run 4 in Table 1).
The notation is the same as in Figure 3.
}

% Figure 10.
\figcaption[fig10.ps]{
Top panel: the logarithmic rate of change
of the cooling radius with time as
a function of $\beta$ for flows in a static potential.
Bottom panel: the logarithmic rate of change of the
dynamical mass accretion rate with time as
a function of $\beta$.
Filled figures correspond to models with $q=4$ and open
figures to $q=1$; The triangles represent values that one
would estimate by applying equation (4) to
ideal observations of the model clusters,
while the squares are the values obtained
by direct measurement from the simulations.
Here, the failure of equation (4) results from
two separate causes.  In the $q=1$ models the cooling flows are
too young for complete cooling in the center, and in the
$q=4$ models the strong mass deposition results in structural
changes that reach far beyond the fiducial cooling radius.
Because the value of $q$ in real clusters is not known, we must
conclude that it is very difficult to measure reliably the rate of change
of
$\dot M$ directly from X-ray images.
}

% Figure 11.
\figcaption[fig1l.ps]{
Steady-state cooling flow solutions for the hot cluster
runs in Figures 2 and 4 obtained using the method of
White \& Sarazin (1987).
The four panels show total number density,
temperature, mass accretion rate and instantaneous isobaric cooling
time
as functions of radius for $q=1$ and $q=4$.
}

% Figure 12.
\figcaption[fig12.ps]{
Same as Figure 11 except for the cool cluster runs of Figures 6 and 8.
}

% Figure 13.
\figcaption[fig13.ps]{
Our simulation of the fiducial run of M90 with a central
galaxy and $q=0$.  The evolution times are $3.3 \Gyr$ (initial
conditions; dotted line),
4.0 (short dash), 7.0 (long dash) and $9.9 \Gyr$ (dot-dash).  A
cold core
has developed in the innermost grid zone by $9.9\Gyr$, preventing
continued
numerical evolution.
}

% Figure 14.
\figcaption[fig14.ps]{
Same as Figure 13 but for $q=1$.  The
dot-dashed and solid lines correspond here to $t=10.0$ and
$13.0 \Gyr$.
}

% Figure 15.
\figcaption[fig15.ps]{
The evolution of cooling radius parameters in our simulation of the
the $q=1$ M90 model (same notation as Figure 3).
The structure in the time-dependence of the mass accretion
rate is due to transient sound waves and shocks resulting from
non-static initial conditions.
Despite these effects, the dynamical and cooling accretion
rates trace each other quite well, indicating that the flow has
reached a steady state.
}

% Figure 16.
\figcaption[fig16.ps]{
Same as Figure 14 but for $q=4$.
}

% Figure 17.
\figcaption[fig17.ps]{
Same as Figure 16 but for $q=4$.
}

\clearpage

%
% table of numerical models
%
\begin{table}[thb]
\caption{\hfil Model Parameters \label{tab:models} \hfil}
\begin{center}
\begin{tabular}{lccccccccc}
\tableline
\tableline
% Column titles.
		 Run                               % 1
 	&        $\sigma_c            $           % 2
 	&        $q                   $           % 3
 	&        $T_{\infty}          $           % 4
 	&        $\beta               $           % 5
 	&        $n_0                 $\tablenotemark{a}           % 6
 	&        $t_{\cool,0}         $\tablenotemark{a}           % 7
 	&        $r_{\cool}           $           % 8
 	&        $\dot M|_{r_{\cool}} $           % 9
 	&        $T|_{r_{\cool}}      $       \cr % 10
% Physical units.
 	&        $(\kms)                 $           % 2
 	&        $                       $           % 3
 	& \hfill $(10^7\K)               $\hfill     % 4
 	&        $                       $           % 5
 	&        $(\den)                 $           % 6
 	&        $(\Gyr)                 $           % 7
 	&        $(\kpc)                 $           % 8
 	&        $(\mdot)                $           % 9
 	& \hfill $(10^7\K)               $\hfill \cr % 10
\tableline
\phn1  &    1054 & 1 &    10.0 & 0.83 & 1.8 &    10.8 &  86 &
278\phd\phn & 6.5 \cr
\phn2  &    1054 & 4 &    10.0 & 0.83 & 3.3 & \phn5.9 &  69 &
244\phd\phn & 7.6 \cr
\phn3  & \phn527 & 1 & \phn3.0 & 0.69 & 1.8 & \phn5.2 &  68 &
\phn65\phd\phn & 2.1 \cr
\phn4  & \phn527 & 4 & \phn3.0 & 0.69 & 3.3 & \phn2.8 &  60 &
\phn93\phd\phn & 2.4 \cr
\phn5  & \phn527 & 1 & \phn1.5 & 1.38 & 1.8 & \phn2.0 &  23 &
\phn\phn1.4 & 1.2 \cr
\phn6  & \phn527 & 4 & \phn1.5 & 1.38 & 3.3 & \phn1.1 &  23 &
\phn\phn2.7 & 1.3 \cr
\phn7  & \phn527 & 1 & \phn2.0 & 1.04 & 1.8 & \phn3.0 &  45 &
\phn10\phd\phn & 1.5 \cr
\phn8  & \phn527 & 4 & \phn2.0 & 1.04 & 3.3 & \phn1.6 &  39 &
\phn15\phd\phn & 1.6 \cr
\phn9  & \phn527 & 1 & \phn5.0 & 0.42 & 1.3 &    10.0 &  64 &
\phn 53\phd\phn & 3.3 \cr
10     & \phn527 & 4 & \phn5.0 & 0.42 & 1.3 &    10.0 &  34 &
\phn18\phd\phn & 3.3 \cr
\tableline
\end{tabular}
\end{center}
\tablenotetext{a}{Values with a subscript of $0$ refer to initial central
values at a radius $r = 0$.}

\end{table}


\clearpage

\begin{references}

\reference{}
Allen, S. W., Fabian, A. C., Johnstone, R. M., Arnaud, K. A., \&
Nulsen, P. E. J. 2000, \mnras, in press (astro-ph/9910188)

\reference{}
Bailey, M. E., \& MacDonald, J.  1981, \mnras, 194, 195

\reference{}
Bertschinger, E. 1988, in Cooling Flows in Clusters and Galaxies,
ed.\ A. C. Fabian, (Dordrecht: Kluwer), 337

\reference{}
Bertschinger, E. 1989, \apj, 340, 666

\reference{}
Binney, J., \& Tremaine, S. 1987, Galactic Dynamics, (Princeton:
Princeton University Press)

\reference{}
Bird, C. M. 1994, \aj, 107, 1637

\reference{}
Bregman, J. N., \& David, L. P. 1988, \apj, 326, 639

\reference{}
Bregman, J. N., \& David, L. P. 1989, \apj, 341, 49

\reference{}
Canizares, C. R., Markert T. H., \& Donahue M. E. 1988, in Cooling
Flows in Clusters and Galaxies, ed.\ A. C. Fabian (Dordrecht: Kluwer), 63

\reference{}
Chevalier, R. A. 1987, \apj, 318, 66

\reference{}
Chevalier, R. A. 1988, \apj, 329, 16

\reference{}
Evrard, A. E. 1990, \apj, 363, 349

\reference{}
Fabian, A. C. 1994, ARA\&A, 32, 277

\reference{}
Gaetz, T., \& Salpeter E. 1983, \apjs, 52, 155

\reference{}
G\'omez, P. L., Loken, C., Burns, J. O., \& Roettinger, K. 2000, preprint

\reference{}
Hawley, J. F., Smarr, L. L., \& Wilson J. R. 1984, \apjs, 55, 211

\reference{}
Jones, C., \& Forman, W. 1984, \apj, 276, 38

\reference{}
Lufkin, E. A., Balbus, S. A., \& Hawley, J. F. 1995, \apj, 446. 529

\reference{}
Lufkin, E. A., \& Hawley, J. F. 1993, \apjs, 88, 569

\reference{}
Markevitch, M., Forman, W. R., Sarazin, C. L., \& Vikhlinin, A. 1998,
\apj, 503, 77

\reference{}
McGlynn, T. A., \& Fabian, A. C. 1984, \mnras, 208, 709

\reference{}
Meiksin, A. 1988, \apj, 334, 59

\reference{}
Meiksin, A. 1990, \apj, 352, 466 (M90)

\reference{}
Navarro, J. F., Frenk, C. S., \& White, S. D. M. 1997, \apj, 490, 493

\reference{}
Peres, C. B., Fabian, A. C., Edge, A. C., Allen, S. W., Johnstone, R. M.,
\& White, D. A. 1998, \mnras, 298, 416

\reference{}
Raymond, J. C., Cox, D. P., \& Smith, B. W. 1976, \apj, 204, 290

\reference{}
Ruppell, H. M., \& Cloutman, L. D. 1975, Los Alamos Nat.\ Lab.\ Rept.\
LA 6149-MS

\reference{}
Sarazin, C. L., \& Graney, C. M. 1991, \apj, 375, 552

\reference{}
Stewart, G. C., Canizares, C. R., Fabian, A. C., \& Nulsen, P. E. J.
1984, \apj, 278, 536

\reference{}
Thomas, P. A., Fabian, A. C., \& Nulsen, P. E. J. 1987, \mnras, 228, 973

\reference{}
White, D. A. 2000, \mnras, in press (astro-ph/9909467)

\reference{}
White, D. A., Jones, C., \& Forman, W. 1997, \mnras, 292, 419

\reference{}
White, R. E. III 1991, \apj, 367, 69

\reference{}
White, R. E. III 1988, in Cooling Flows in Clusters and Galaxies,
ed.\ A. C. Fabian, (Dordrecht: Kluwer), 343

\reference{}
White, R. E. III, \& Sarazin, C. L. 1987, \apj, 318, 629 (WS)

\reference{}
Wise, M. W. \& Sarazin, C. L. 1993, \apj, 415, 58

\end{references}
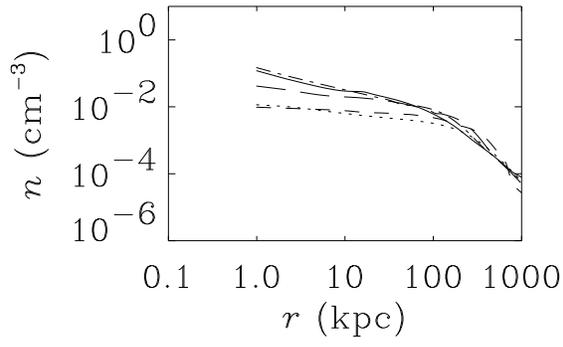
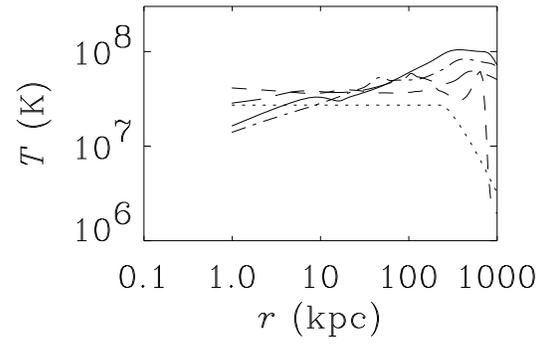
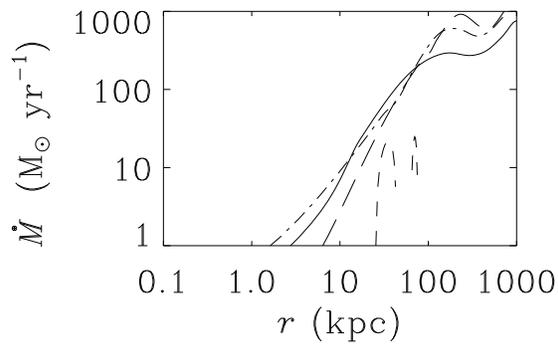
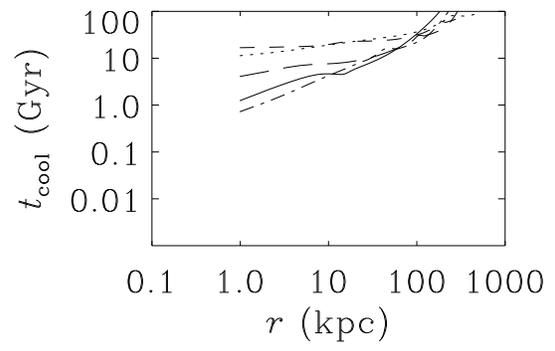
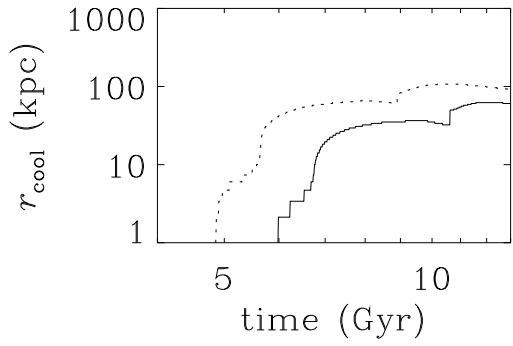
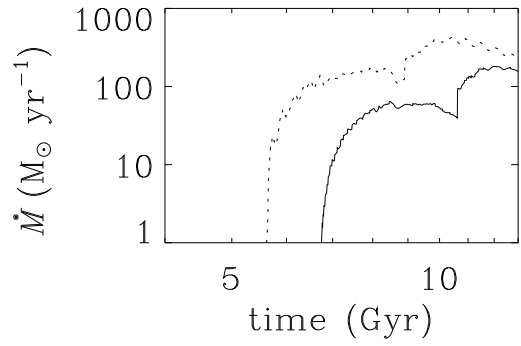
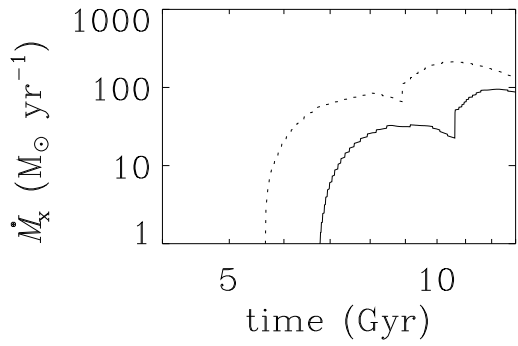
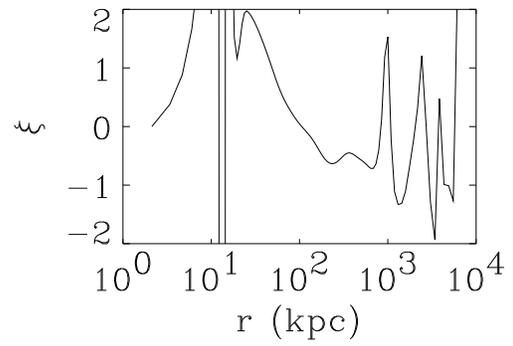
\end{document}